\documentclass[aps,pra,twocolumn,amsmath,amssymb,nofootinbib,superscriptaddress]{revtex4}
\usepackage[english]{babel}
\usepackage{latexsym}
\usepackage{graphics}
\usepackage{epsfig}
\usepackage{color}
\usepackage{bm}
\usepackage{amsmath}
\usepackage{amssymb}
\usepackage{array,multirow}

\usepackage{stackengine}
\stackMath
\newcommand\tenq[2][1]{%
 \def\useanchorwidth{T}%
  \ifnum#1>1%
    \stackon[0pt]{\tenq[\numexpr#1-1\relax]{#2}}{\scriptscriptstyle\sim}%
  \else%
    \stackon[1pt]{#2}{\scriptscriptstyle\sim}%
  \fi%
}

\usepackage[normalem]{ulem}
\usepackage{color}

\newcommand{\be}{\begin{equation}}
\newcommand{\ee}{\end{equation}}
\newcommand{\Rvec}{\ensuremath{\boldsymbol{R}}}
\newcommand{\etavec}{\ensuremath{\boldsymbol{\eta}}}

\begin{document}

\title{Mass-imbalanced fermionic mixture in a harmonic trap}
\author{B. Bazak}
\affiliation{IPNO, CNRS/IN2P3, Univ. Paris-Sud, Universit\'e Paris-Saclay,
  F-91406, Orsay, France }

\date{\today}

\begin{abstract}
  The mass-imbalanced fermionic mixture is studied, where $N\le5$ identical
  fermions interact resonantly with an impurity, a distinguishable atom.
  The shell structure is explored, and the physics of a dynamic light-impurity
  is shown to be different from that of the static heavy-impurity case.
  The energies in a harmonic trap at unitarity are calculated 
  and extrapolated to the zero-range limit.
  In doing so, the scale factor of the ground state, as well as of a few
  excited states, is calculated.
  In the $2 \le N \le 4$ systems, pure $(N+1)$ Efimov states exist for large
  enough mass ratio.
  However, no sign for a six-body Efimov state in the $(5+1)$ system is found
  in the mass ratio explored, $M/m \le 12$.
\end{abstract}

\maketitle

\section{INTRODUCTION}
The system of $N$ identical fermions interacting resonantly with a
distinguishable atom exhibits a rich and interesting physics, including
universal phenomena and the celebrated Efimov physics.
For a recent review see, e.g., Ref.~\cite{NaiEnd16}.

An important parameter here is the ratio of the impurity mass $m$ and the
identical fermions mass $M$. 
In the ultracold limit the interaction between identical fermions can
be neglected, and therefore in the heavy impurity case $m \gg M$ the problem
is decoupled to $N$ independent fermions interacting with a static impurity.
The opposite limit, where $m \ll M$, corresponds to a
dynamic impurity which induces interaction between the identical fermions.

The simplest non trivial example is the $(2+1)$ system, composed of two
identical fermions of mass $M$ and a distinguishable atom of mass $m$,
where different particles have zero-range resonant interaction
while identical particles do not interact.
Efimov has shown that when the mass ratio $\alpha=M/m$ is larger than the
critical value $\alpha_c=13.607$, an infinite tower of trimers with
angular momentum and parity $L^\pi=1^-$ is produced \cite{Efimov1973}.
The $n$-th trimer energy is $E_n=E_0e^{-2\pi n/|s|}$, where $E_0$ is the
trimer ground-state energy. The scale factor $s=s(\alpha)$
is a function of the mass ratio and vanishes at the Efimov threshold
$s(\alpha_c)=0$.

In the non-Efimovian regime $\alpha<\alpha_c$ the scale
factor characterizes the short-distance (and large momenta)
behavior of a universal trimer, which exists for $8.173<\alpha<\alpha_c$
for finite positive scattering length \cite{KarMal07}.

The physical interpretation of the scale factor can be understood
from the adiabatic hyperspherical formalism \cite{Mac68}.
To see that, one rearranges the relative coordinates into the hyperradius
$\rho$, the only coordinate with a dimension of length,
and $3N-1$ hyperangles.
Here $\rho \propto \sqrt{m r^2+M\sum_{i=1}^{N}R_i^2}$,
where ${\bf r}$ (${\bf R_i}$) is the position of the
distinguishable (identical) atom in the center-of-mass frame.
At small $\rho$, where $E$ and $1/a$ can be neglected, the hyperradial motion
separates from hyperangular degrees of freedom and is governed by 
\begin{equation}\label{Schr}
  \left[-\frac{\partial^2}{\partial \rho^2}-\frac{3N-1}{\rho}
    \frac{\partial}{\partial \rho}+\frac{s^2-(3N/2-1)^2}{\rho^2}\right]
  \Psi(\rho)=0,
\end{equation}
where $s^2$ is the hyperangular eigenvalue.
The general solution of Eq.~(\ref{Schr}) is a linear combination of
$\Psi_+(\rho)\propto \rho^{-3N/2+1+s}$ and $\Psi_-(\rho)\propto \rho^{-3N/2+1-s}$.
The case $s^2<0$ ($s=is_0$) corresponds to the Efimovian regime, where this
linear combination is an oscillating function, and a three-body parameter
is required to fix the relative phase of $\Psi_+$ and $\Psi_-$.
The non-Efimovian regime appears for $s^2>0$ ($s>0$) where,
far from few-body resonances, $\Psi(\rho)$ is dominated by $\Psi_+(\rho)$.

Interestingly, the same factor determines the energy of the trapped
system at unitarity \cite {Tan04,WerCas06}, namely,
\be\label{trap}
E=\hbar \omega (s+2n+1),
\ee 
where $\omega$ is the trapping frequency, taken to be identical for all
particles, $n$ is a non-negative integer and the center-of-mass zero-point
energy is omitted. 
This is because the trapping potential is involved only in the hyperradial
equation, while $s$ is determined by the hyperangular equation which
is identical in free space and in a trap.
For a recent review of the trapped few-body problem, see Ref.~\cite{Blu12}.

Following Efimov, the mass-imbalanced (2+1) system has attracted wide attention
(see, e.g., Refs.~\cite{Efimov1973,KarMal07,Pet03,Fon79,PetSalShl04,NisSonTan08,
  LevTieWal09, RitMehGre10,MatParHus11,HelHam11,EndNaiUed11,LevPet11,CasTig11,
  Safavi2013,KarMal16,EndCas16}).
The scale factor of the (2+1) system was first calculated for the equal-mass
case to be $s(1) = 1.7727$ for the $1^-$ ground state and
$s(1) = 2.1662$ for the $0^+$ excited state \cite{PetSalShl04}.
Later, the method was generalized to include any angular momentum and
mass ratio \cite{RitMehGre10}.
The $1^-$ trimer energy crosses the dimer+atom energy in a trap at
$\alpha=8.6186$ \cite{Pet03}.
An ultracold mixture of $^6$Li and $^{40}$K
($\alpha \approx 6.4$) was realized experimentally, and
a strong atom-dimer attraction was observed.
This attraction was interpreted as $p$-wave interaction
between two heavy particles induced by the light atom \cite{Rudi14}.

The trend of moving from a non-Efimovian universal state to an Efimovian state
with the same symmetry as the mass ratio increases was discovered also in the
$(3+1)$ and $(4+1)$ systems \cite{CasMorPri10,Blu12b,BazPet17}.

The mass-imbalanced $(3+1)$ system has been the subject of a few studies
\cite{CasMorPri10,Blu12b,BazPet17,BluDai10,EndCas16}.
Here a tower of $1^+$ Efimovian tetramers exists above $\alpha_c=13.384$
\cite{CasMorPri10}, and a universal non-Efimovian $1^+$ tetramer is bound in
free space for $8.862<\alpha<\alpha_c$ \cite{Blu12b,BazPet17}.
The scale factor of the tetramer ground state has been calculated for a few
mass ratios \cite{BluDai10}, while that of excited states is known only for
the equal-mass case \cite{RakDaiBlu12}.
The tetramer energy crosses the trimer+atom energy in a trap at
$\alpha=8.918$ \cite{BazPet17}.

The mass-imbalanced $(4+1)$ system was studied in
Refs.~\cite{BluDai10,BazPet17}.
A tower of $0^-$ Efimovian pentamers exists above $\alpha_c=13.279$,
while a universal $0^-$ pentamer is bound in free space
for $9.672<\alpha<\alpha_c$  \cite{BazPet17}.
Here the scale factor is known for 
equal mass \cite{RakDaiBlu12}, when the pentamer is bound in free space
\cite{BazPet17} and for few other mass ratios \cite{BluDai10}.
The pentamer energy crosses the tetramer+atom energy in a trap at $\alpha=9.41$
\cite{BazPet17}.

The ground-state properties of the $(N+1)$ systems are summarized in
Table~\ref{tbl:Thresholds}.

\begin{table}
\begin{center}
  \caption{The ground-state properties of mass-imbalanced $(N+1)$ fermionic
    mixtures, for $N\le5$.
    Shown are the angular momentum and parity of the state,
    the mass ratio where it crosses the threshold of the system with one
    particle less in free space and in a harmonic trap,
    and the mass ratio where Efimov physics emerges.
    See text for references.
    \label{tbl:Thresholds}}
\vspace{0.3cm}
{\renewcommand{\arraystretch}{1.25}%
\begin{tabular}
{c@{\hspace{5mm}} c@{\hspace{5mm}}  c@{\hspace{5mm}} c@{\hspace{5mm}} c}
\hline\hline 
System & $L^\pi$ & Free crossing & Trap crossing & Efimov \\
\hline
2+1 & $1^-$ & 8.173 & 8.619 & 13.607 \\
3+1 & $1^+$ & 8.862 & 8.918 & 13.384 \\
4+1 & $0^-$ & 9.672 & 9.41  & 13.279 \\
5+1 & $0^-$ \\
\hline\hline
\end{tabular}}
\end{center}
\end{table}

Very little is known about the $(5+1)$ system.
A simplified model explains the similar trends in the 
$(2+1)$, $(3+1)$, and $(4+1)$ systems as populating a $p$ shell atom
by atom. The $(5+1)$ system, therefore, should be different, since the
$p$ shell is now full and the additional atom has to open a new shell
\cite{BazPet17}.
Intriguing open questions are thus the following:
is there a non-Efimovian universal bound hexamer
and does the six-body Efimov effect exist?

The extrapolation toward the case of fermionic polaron,
corresponding to the $N \gg 1$ case, is of special interest.
As a step in this direction the shell structure of the few-body
systems is studied here.
In contrast to the static heavy-impurity case, it is shown 
that non perturbative physics arise in the dynamic
light-impurity case.

The goal of this work is to study the scale factor, or equivalently the
energy in a trap, of the $(N+1)$ ($N \le 5$) fermionic mixtures
few lowest states, and to identify their properties.
Calculation are done for a wide range of mass ratios, 
from the static-impurity limit $m \gg M$ to the 
dynamic-impurity limit $m \ll M$.

A convenient way to describe the system is the
Skorniakov and Ter-Martirosian (STM) integral equation
\cite{STM,MorCasPri11}, which deals directly with zero-range interaction by
applying the Bethe-Peierls boundary condition when two different particles
approach each other.
One has to solve an integral equation in $3(N-1)$ dimensions, 
but utilizing the system symmetries the number of dimensions 
can be reduced further.

For $N=2$, the STM equation for the scale factor is reduced to a
transcendental equation which can be easily solved.
For $N=3$, it can be reduced to two dimensions,
allowing the solution on a grid \cite{CasMorPri10}.
For $N=4$, however, a five-dimensional equation 
makes a grid-based approach challenging if possible at all.
A method based on a Monte-Carlo process to solve the STM equation was
developed for this case in Ref.~\cite{BazPet17}.
However, this method is limited to bound systems and therefore
cannot be used to calculate the scale factor for all mass ratios.
In addition, as a fermionic Monte-Carlo method it might suffer from
a sign problem if the wave function has radial nodes.

Thus we take here another approach.
We solve the Schr\"{o}dinger equation for 
the trapped system with \emph{finite}-range interspecies potential
and then extrapolate to the zero-range limit.
A similar method was applied in Refs.~\cite{BluDai10,RakDaiBlu12}.

Using this method we calculate the scale factor for $0 \le \alpha \le 12$
for the ground state, as well as for a few lowest excited states, of the
$(N+1)$ fermionic system up to $N \le 5$.
We set a simple model to understand the shell structure for the static-impurity
case, and explore the effects of the dynamic impurity as the mass ratio
increases.

We find that no $(5+1)$ Efimov states exist for $\alpha \le 12$.
As the mass ratio increases, finite-range corrections become
significant and the extrapolation to the zero-range limit cannot be trusted
anymore. A further study is therefore needed to explore such states for
larger mass ratios, $12 < \alpha < 13.279$.

%================
\section{METHODS}
%================
As we have explained, the zero-range limit is not directly used here; instead, 
a series of calculations with a finite-range potential with decreasing range
is used to extrapolate the zero-range limit.

The Hamiltonian of the $(N+1)$ system is
\be
H = T + U + V,
\ee
where $T$ is the internal kinetic energy
and $U$ is the confining harmonic potential. 
Here, $V$ is the interspecies attractive interaction,
taken of the form
\be
V=-V_0 \sum_{i=1}^N \exp\left(-\frac{({\bf r}-{\bf R}_i)^2}{2R_0^2}\right),
\ee
where $V_0>0$ is the potential strength and $R_0$ is its range.
We seek the limit of $R_0 \longrightarrow 0$
while $V_0$ is tuned to keep the two-body system at unitarity.

To solve the few-body problem, we use the stochastic variational method (SVM)
\cite{SuzVar98}. The wave function is expanded in an over-complete basis
of correlated Gaussians, where the basis functions are chosen in a stochastic
way utilizing the variational principle. The energies and the corresponding wave
functions can be found then by solving a generalized eigenvalue problem.

The basis functions are chosen to have the necessary permutational symmetry,
parity $\pi$, and angular momentum $L$ and its projection $M$,
\be
\phi^\pi_{LM}(A,u;\eta) = \hat {\mathcal A} e^{-\frac{1}{2}\eta^T A \eta}\,
\theta^\pi_{LM}(u;\eta)
\ee
where $\eta \equiv \{\etavec_1,\ldots,\etavec_N\}$ is a set of $N$ Jacobi
coordinates, $\hat {\mathcal A}$ is the appropriate anti-symmetrization
operator, 
$A$ is an $N \times N$ real, symmetric, and positive definite matrix,
and $\theta^\pi_{LM}(u;\eta)$ is the angular part.
The $N(N+1)/2$ real numbers defining $A$ are optimized in a stochastic way
such as the energy is minimized.
Spin and isospin functions can be introduced but are not needed here.

The angular part is characterized by the global vector
representation \cite{VarSuz95,SuzUsu00}.
For a natural parity $\pi=(-1)^L$ it is
\be
\theta^\pi_{LM}(u;\eta) = \mathcal Y_{LM}(\mathbf v),
\ee
where $\mathcal Y_{LM}$ is the regular solid harmonic
and $\mathbf{v}=u^T\eta$ is a global vector, whose 
elements are also optimized in a stochastic way.

To get the unnatural parity $\pi=(-1)^{L+1}$ for $L>0$ one has to couple two
global vectors,
\be
\theta^\pi_{LM}(u;\eta) =
\left[\mathcal Y_L(\mathbf v_1) \otimes \mathcal Y_1(\mathbf v_2)\right]_{LM},
\ee
while three global vectors are needed to get the $0^-$ symmetry, 
\be
\theta^-_{00}(u;\eta) =
\left[\left[\mathcal Y_1(\mathbf v_1) \otimes \mathcal Y_1(\mathbf v_2)\right]_1
  \otimes \mathcal Y_1(\mathbf v_3)\right]_{00}.
\ee
The overlap of such basis functions, as well as the matrix elements of the
Hamiltonian, are known analytically
\cite{VarSuz95,SuzVar98,SuzUsu00,RakDaiBlu12,BazEliKol16}.

For a given number of particles, angular momentum, and parity,
the ground-state energy is calculated for various potential ranges. From
these energies, the zero-range limit is extrapolated.

\begin{figure}
\vskip 0 pt \includegraphics[clip,width=1\columnwidth]{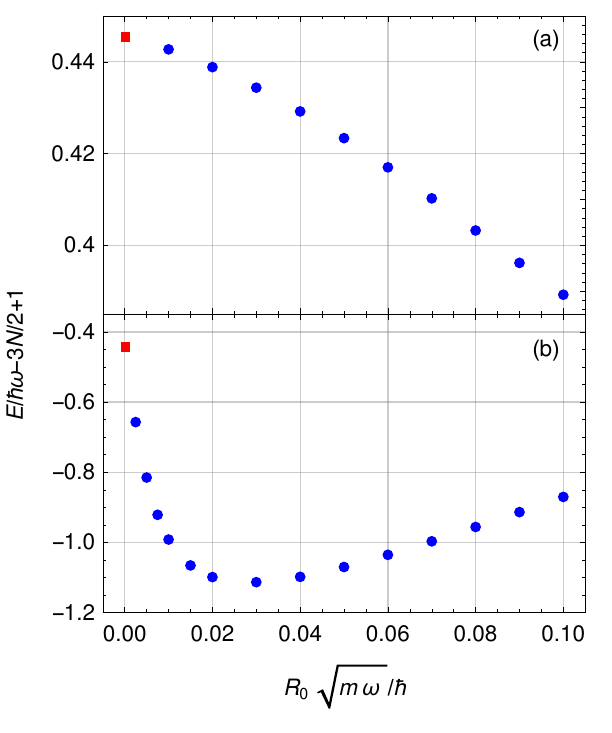}
\caption{
  Convergence of finite-range potentials toward the zero-range limit 
  $R_0 \longrightarrow 0$ for the $(2+1)$ ground state.
  (a) $\alpha=4$, away from the Efimovian limit.
  (b) $\alpha=12$, near the Efimovian limit. 
  The zero-range result (red square) is the exact solution of Eq. (\ref{ZR}).
}
\label{Fig:Convergence}
\end{figure}

Typical results for the (2+1) $1^-$ ground state are shown in
Fig.~\ref{Fig:Convergence}, where results calculated from finite-range
potentials are compared to the zero-range results.
The radius of convergence for the extrapolation is shown to be much larger
for $\alpha=4$ than for $\alpha=12$.
In the latter case, close to the Efimovian limit,
the extrapolated value will be completely off if one uses, say,
results with $R_0>0.03 \sqrt{\hbar^2/m \omega}$ \cite{BluDai10}. 

To estimate the extrapolation uncertainty, we fit the results with
a few shortest $R_0$ with linear and parabolic curves
and account for their differences. The error due to the finite basis set
becomes significant for $N>3$ and is also considered.

Taking the potential range to be smaller, the numerical calculation
becomes harder. 
Therefore close to the Efimovian limit, where finite-range corrections become
significant, the extrapolations can not be trusted anymore.
To correctly treat this region one should use a method dealing with the
zero-range limit directly. For example, one would like to solve the
STM equation using a diffusion Monte-Carlo (DMC)-like approach \cite{BazPet17}.
This task is left for future work. 

%================
\section{RESULTS}
%================~~~~~~~~~~~~~~~~~~
\subsection{The $\alpha = 0$ limit}
%~~~~~~~~~~~~~~~~~~~~~~~~~~~~~~~~~~
We start to analyze the $\alpha = 0$ limit, where the impurity 
is infinitely heavy and therefore static.
This case reduces to the problem of $N$ 
trapped fermions scattering on a zero-range potential at
the trap center.
The analytic solution for the two-body problem is known \cite{BusEngRza98},
giving at unitarity an energy shift of $-\,\hbar\omega$ for the $s$ shell
with respect to the non interacting case.
The quantum numbers characterizing a shell are the radial number $n$ and
the angular momentum $l$ and its projection; its energy is given by
\be
E_{nl}=\hbar \omega (2n+l-\delta_{l,0}+3/2),
\ee
and the energy of the $(N+1)$ system is just a sum of $N$ single-particle
energies.
To ease comparison between clusters with different particle numbers,
the zero-point energy $\hbar \omega \,3N/2$ is subtracted.
Energy is measured in units of $\hbar \omega$
and with respect to the dimer energy, i.e.
\be
\epsilon = E/\hbar \omega-3N/2+1.
\ee

Only interacting states, i.e., those states which have 
an atom in an $s$ shell, are considered.

Applying the fermionic symmetry, the spectrum and properties of the $(N+1)$
systems can be calculated.
Table~\ref{tbl:GroundState} summarizes
the ground-state properties of the $(N+1)$ systems.
For completeness, the properties of the two lowest excited states
are also tabulated in the Appendix. Here and thereafter we ignore the
trivial $2L+1$ degeneracy due to different total angular momentum projections.

\begin{table}
\begin{center}
  \caption{The ground-state properties in the static-impurity limit,
    $\alpha=0$.
    Shown are the energy, the angular momentum, the parity, and the shell
    configuration for the $(N+1)$ mixtures.
    \label{tbl:GroundState}}
\vspace{0.3cm}
{\renewcommand{\arraystretch}{1.25}%
\begin{tabular}
{c@{\hspace{5mm}} c@{\hspace{5mm}}  c@{\hspace{5mm}} c@{\hspace{5mm}} c}
\hline\hline 
System & $\epsilon$ & $\pi$ & $L$ & Configuration \\
\hline
1+1 & 0 & + & 0 & $0s$ \\
\hline
\multirow{2}{*}{2+1} & \multirow{2}{*}{1} & + & 0 & $0s\,1s$ \\
\cline{3-5}
& & -- & 1 & $0s\,0p$ \\
\hline
\multirow{2}{*}{3+1} & \multirow{2}{*}{2} & + & 1 & $0s\,0p^2$ \\
\cline{3-5}
& & -- & 1 & $0s\,1s\,0p$ \\
\hline
\multirow{2}{*}{4+1} &\multirow{2}{*}{3} & + & 1 & $0s\,1s\,0p^2$ \\
\cline{3-5}
& & -- & 0 & $0s\,0p^3$ \\
\hline
5+1 & 4 & -- & 0 & $0s\,1s\,0p^3$ \\
\hline\hline
\end{tabular}}
\end{center}
\end{table}

%~~~~~~~~~~~~~~~~~~~~~~~~~~~
\subsection {The (2+1) case}
%~~~~~~~~~~~~~~~~~~~~~~~~~~~
We move now to the general mass-imbalanced case and start with 
two identical fermions interacting with a distinguishable atom.

For the natural parity case, the scale factor $s$ corresponding to
a total angular momentum $L$ is
the solution of a transcendental equation,
\be\label{ZR}
\frac{2}{\Gamma(a-1/2)\Gamma(b-1/2)}+
\frac{\left(-\gamma\right)^L}{\sqrt \pi \Gamma (c)} 
\,_2F_1\left(a,b;c;\gamma^2\right)=0
\ee
where $a=1+(L-s)/2$, $b=1+(L+s)/2$, $c=L+3/2$, 
$_2F_1$ is the hypergeometric function, and $\gamma=\alpha/(\alpha+1)$
\cite{RitMehGre10}.

Unnatural parity means here that both identical fermions
are excited to $l>0$ shell, resulting in a non interacting case that will be
ignored here. 

For $\alpha=0$ the ground state has two degenerate states, $1^-$ and $0^+$,
where in the first case the additional atom populates a $p$ shell while in the
latter it sits in an excited $s$ shell.
The energy degeneracy is lifted for $\alpha>0$, where the dynamic impurity
induces interaction between the identical fermions,
which is attractive (repulsive) for an odd (even) angular momentum.
Hence, the $1^-$ state becomes the ground state.

This behavior can be understood in the Born-Oppenheimer (BO)
approximation, which holds for $\alpha \gg 1$ \cite{Fon79}.
Utilizing the mass difference,
the distance between heavy particles $\Rvec=\Rvec_1-\Rvec_2$
can be treated as a parameter in the light-particle equation,
which becomes simply the double-well potential problem,
with the known eigenvalues $\epsilon_\pm(R)$.
In the heavy-particle equation, $\epsilon_\pm(R)$ has the meaning of an
effective potential and is attractive or repulsive, depending on the parity.
Applying the fermionic symmetry for heavy particles' permutation, the
effective potential for odd-$L$ states is found to be attractive and
goes like $-1/mR^2$ for $R \ll a$, while the effective potential for
even-$L$ states is repulsive.

For the attractive channel, the mass ratio governs the competition between
the centrifugal barrier $\propto L(L+1)/MR^2$ and the effective attraction. 
Increasing $\alpha$ tips the scales in favor of the attraction;
hence the trimer energy decreases.
In a trap the trimer energy crosses the dimer+atom energy
($\epsilon=0$ in our conventions)
for $\alpha$ slightly larger than needed in free space.
Increasing $\alpha$ further the effective interaction becomes purely
attractive and the system becomes Efimovian.
In the $(2+1)$ system, the $1^-$ symmetry is the only symmetry where this
phenomenon occurs.

To benchmark our method, we calculate the unitary $(2+1)$ trapped system
energy by extrapolating finite-range results to the zero-range limit.
The scale factor can be easily calculated from Eq. (\ref{ZR}) and is connected
to the energy in a trap by Eq. (\ref{trap}), giving here (for $n=0$)
$s=\epsilon+1$.
Hence, the Efimovian limit $s=0$ corresponds here to $\epsilon=-1$.
Our results are plotted in Fig.~\ref{Fig:Trimer}, showing
a nice agreement with the solutions of Eq. (\ref{ZR}).
The limit of $\alpha=0$ from Tables \ref{tbl:GroundState}, 
\ref{tbl:ExcitedState} and \ref{tbl:2ExcitedState} is also reproduced.

\begin{center}
\begin{figure}
\vskip 0 pt \includegraphics[clip,width=1\columnwidth]{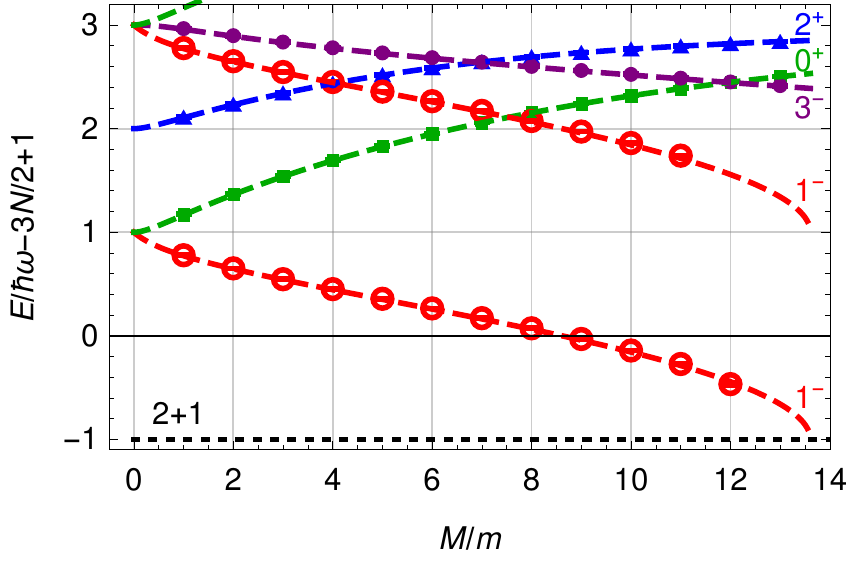}
\caption{
  The energy of the unitary $(2+1)$ trapped system is shown as a function of
  the mass ratio for a few lowest states.
  Symbols are the zero-range extrapolation from finite-range potentials,
  and dashed curves are the zero-range results calculated from  Eq. (\ref{ZR}).
  The Efimovian limit $s=0$ is the dotted horizontal line, which the lowest
  $1^-$ curve hits at $M/m=13.607$.
}
\label{Fig:Trimer}
\end{figure}
\par\end{center}

Note that in a trap, each solution of Eq. (\ref{ZR}) starts a ladder of
solutions, corresponding to hyperradial excitations and giving an additional 
$2\hbar\omega$ for each hyperradial node.
The first excited state of the $1^-$ symmetry is also shown
in Fig.~\ref{Fig:Trimer}.

%~~~~~~~~~~~~~~~~~~~~~~~~~~~
\subsection {The (3+1) case}
%~~~~~~~~~~~~~~~~~~~~~~~~~~~

We now add another identical particle and move to the $(3+1)$ system.
For $\alpha=0$, the ground state has two degenerate states, $1^+$ and $1^-$,
both with $\epsilon=2$.
These states have different atomic configurations: while in the $1^+$
state the additional atom sits in a $p$ shell, the $1^-$ state 
corresponds to atom-trimer $s$-wave scattering. $d$-wave atom-trimer scattering
states, corresponding to $1^-$, $2^-$, and $3^-$ symmetries,
have higher energy in this limit, $\epsilon=3$.

The energy degeneracy is lifted for $\alpha>0$, where
the $1^+$ state energy becomes lower than the $1^-$ state energy, in qualitative
agreement with the BO picture where the interaction induced by the impurity is
attractive in a $p$ wave and repulsive in an $s$ wave.

For a larger mass ratio, the $1^+$ state becomes bound in free space,
then crosses the trimer+atom threshold in a trap,
and eventually reaches the Efimov threshold, corresponding here
to $\epsilon=-2.5$.
States of other symmetries, nevertheless, does not reach the Efimov limit for
any mass ratio smaller than the $(2+1)$ Efimov threshold \cite{CasMorPri10}.

\begin{figure}
\vskip 0 pt \includegraphics[clip,width=1\columnwidth]{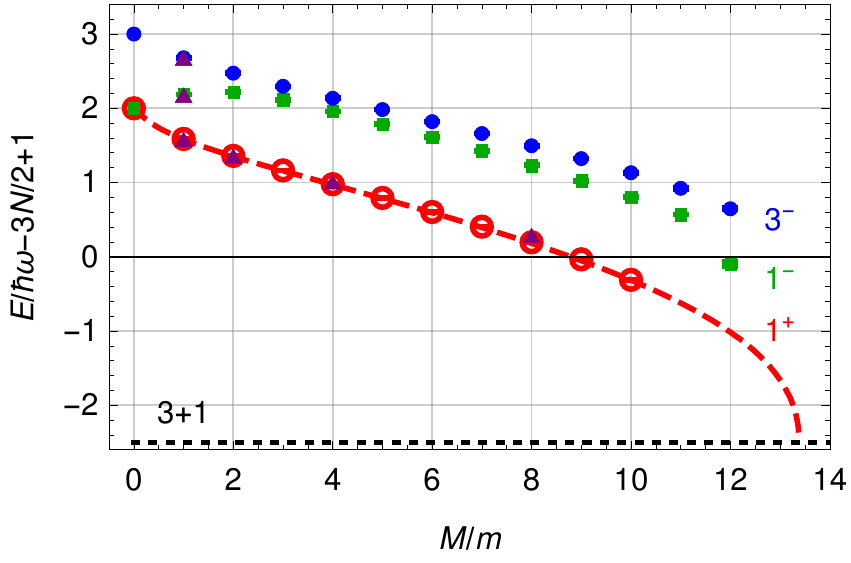}
\caption{
  The energy of the unitary $(3+1)$ trapped system is shown as a function of
  the mass ratio for a few lowest states.
  Symbols are the zero-range extrapolation from finite-range potentials,
  and the dashed curve is the zero-range result of Ref.~\cite{BazPet17}.
  The results of Refs.~\cite{BluDai10,RakDaiBlu12} are shown as purple
  triangles.
  The Efimovian limit $s=0$ is the dotted horizontal line, which the $1^+$
  curve approaches at $M/m=13.384$.
}
\label{Fig:Tetramer}
\end{figure}

The $1^+$ ground-state scale factor has been calculated in Ref.~\cite{BazPet17}
using a grid-based method, similar to that of Ref.~\cite{CasMorPri10}.
That method is more accurate than our current method and can be used up to,
and even beyond, the Efimov limit.
For a benchmark, we compare in Fig.~\ref{Fig:Tetramer} the results of 
both methods, which are in nice agreement.
The $\alpha=0$ limit from Table~\ref{tbl:GroundState} is also reproduced.
For this symmetry the calculations for $\alpha>10$ become sensitive,
signing a non universal resonance, identified in Ref.~\cite{BluDai10} to occur
at $\alpha=10.4(2)$ for a Gaussian interaction.

The scale factor of the $1^-$ lowest excited state has been calculated   
for an equal-mass system only \cite{RakDaiBlu12}.
Our results are tabulated in Table~\ref{tbl:TetramerX} and shown in
Fig.~\ref{Fig:Tetramer}, agreeing well with the $\alpha=0$ limit and with
the $\alpha=1$ result of Ref.~\cite{RakDaiBlu12}.

\begin{figure}
\vskip 0 pt \includegraphics[clip,width=1\columnwidth]{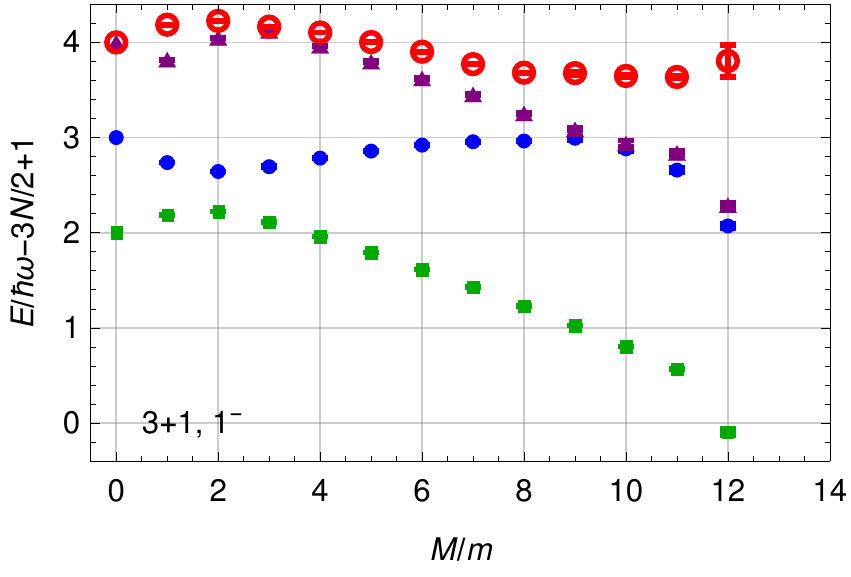}
\caption{
  The energy of the unitary $(3+1)$ trapped system is shown as a function of
  the mass ratio, for a few lowest $1^-$ states.  
}
\label{Fig:TetramerX}
\end{figure}

The bending in the $1^-$ energy around $\alpha=2$ is to be understood as
level repulsion with an excited $1^-$ state.
To make this point clear, the energies of a few lowest $1^-$ states are shown
in Fig.~\ref{Fig:TetramerX}.
The atomic configurations for $\alpha=0$ are the following.
The state with $\epsilon=2$ corresponds to
the configuration $0s\,0p\,1s$, i.e. an atom-trimer $s$-wave state, while
for $\epsilon=3$ it is $0s\,0p\,0d$, i.e. an atom-trimer $d$-wave state.
A clear avoided crossing between these states is seen around $\alpha=2$.

Note, however, that the crossing of levels with different quantum numbers is
allowed.
States with different hyperradial quantum number $n$ can therefore cross, 
and are also shown in Fig.~\ref{Fig:TetramerX}.

\begin{table}
\begin{center}
  \caption{The energies of the trapped tetramer lowest $1^-$ state.
    \label{tbl:TetramerX}}
\vspace{0.3cm}
{\renewcommand{\arraystretch}{1.25}%
\begin{tabular}
{c@{\hspace{5mm}} c@{\hspace{5mm}} c@{\hspace{5mm}} c@{\hspace{5mm}} c}
\hline\hline
$M/m$ & This work & Ref.~\cite{RakDaiBlu12} & $M/m$ & This work \\
\hline
0  & 2        &          & 6  & 1.613(1) \\
1  & 2.183(2) & 2.177(4) & 7  & 1.428(1) \\
2  & 2.221(2) &          & 8  & 1.232(1) \\
3  & 2.115(2) &          & 9  & 1.024(1) \\
4  & 1.959(1) &          & 10 & 0.805(2) \\
5  & 1.791(1) &          & 11 & 0.569(3) \\
\hline\hline
\end{tabular}}
\end{center}
\end{table}

The next state, with $3^-$ symmetry, is also shown in Fig.~\ref{Fig:Tetramer}.
It moves closer to the $1^-$ state as the mass ratio increases.
Since the lowest $1^-$ for large $\alpha$ is dominated by a $d$-wave
atom-trimer state, like the $3^-$ state, this similarity makes sense.
As we show later, this phenomena also exists, 
and is even stronger, for larger $N$.

%~~~~~~~~~~~~~~~~~~~~~~~~~~~
\subsection {The (4+1) case}
%~~~~~~~~~~~~~~~~~~~~~~~~~~~

Adding another identical particle, we now consider the $(4+1)$ system.

For $\alpha=0$, two states are degenerate at $\epsilon=3$, with symmetries 
$0^-$ and $1^+$.
In the $0^-$ state the additional atom populates the last place in
the $p$ shell,
while the $1^+$ state corresponds to atom-tetramer $s$-wave scattering.
The degeneracy is lifted for $\alpha>0$, where
the $0^-$ state energy becomes lower than the $1^+$ energy.
For larger mass ratios, the $0^-$ state
crosses the tetramer+atom energy in a trap,
becomes bound in free space, 
and eventually reaches the Efimov threshold, corresponding here to
$\epsilon=-4$ \cite{BazPet17}.

\begin{figure}
\vskip 0 pt \includegraphics[clip,width=1\columnwidth]{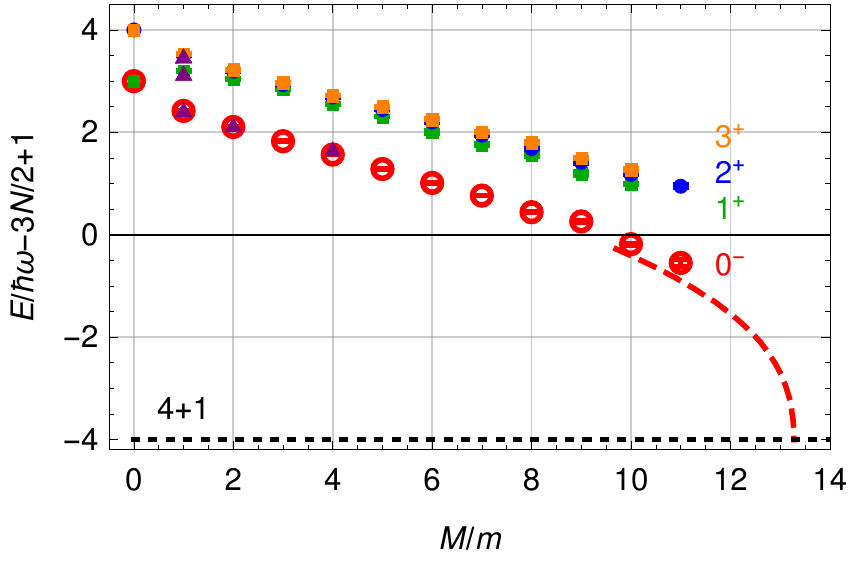}
\caption{
  The energy of the unitary $(4+1)$ trapped system is shown as a function of
  the mass ratio for a few lowest states.
  Symbols are the zero-range extrapolation from finite-range potentials,
  and the dashed curve is the zero-range result of Ref.~\cite{BazPet17}.
  The results of Refs.~\cite{BluDai10,RakDaiBlu12} are shown as purple
  triangles.
  The Efimovian limit $s=0$ is the dotted horizontal line, which the $0^-$
  curve approaches at $M/m=13.279$.
}
\label{Fig:Pentamer}
\end{figure}

The $0^-$ scale factor has been calculated for a few mass ratios using
finite-range models \cite{BluDai10}. 
For $\alpha>9.672$, when the pentamer is bound in free space, it was
calculated by fitting the wave-function high-momentum tail to
$F(Q)\propto Q^{-3N/2+1-s}$, where $Q$ is the hypermomentum conjugate to the
the hyperradius $\rho$ 
and $F$ is the momentum-space wave-function calculated in the STM-DMC method
\cite{BazPet17}.
Our results are tabulated in Table~\ref{tbl:Pentamer} and shown
in Fig.~\ref{Fig:Pentamer}. 

\begin{table}
\begin{center}
  \caption{The energies of the trapped pentamer $0^-$ state for various
    mass ratios.
    \label{tbl:Pentamer}}
\vspace{0.3cm}
{\renewcommand{\arraystretch}{1.25}%
\begin{tabular}
{c@{\hspace{2mm}} c@{\hspace{4mm}} c@{\hspace{3mm}} c@{\hspace{2mm}} c@{\hspace{4mm}} c@{\hspace{4mm}} c}
\hline\hline
$M/m$ & This work & Ref.~\cite{BluDai10} & $M/m$ & This work & Ref.~\cite{BazPet17} \\
\hline
0  & 3       &      &  6 & 1.01(1) \\
1  & 2.42(1) & 2.45 &  7 & 0.77(1) \\
2  & 2.11(1) & 2.15 &  8 & 0.44(1) \\
3  & 1.83(1) &      &  9 & 0.26(3) \\
4  & 1.57(1) & 1.68 & 10 & -0.2(1) & -0.41(1) \\
5  & 1.28(1) &      & 11 & -0.5(1) & -0.90(1) \\
\hline\hline
\end{tabular}}
\end{center}
\end{table}

The $1^+$ scale factor has been calculated only for the equal-mass
case \cite{RakDaiBlu12}.
Our results are tabulated in Table~\ref{tbl:PentamerX} and shown
in Fig.~\ref{Fig:Pentamer}.
Since for large mass ratio the zero-range extrapolation is not conclusive,
we cannot work close to the Efimov threshold. However, no sign
for an Efimov state with any symmetry other than $0^-$ is found in the
explored mass ratios. 

\begin{table}
\begin{center}
  \caption{The energies of the trapped pentamer $1^+$ state for various
    mass ratios.
    \label{tbl:PentamerX}}
\vspace{0.3cm}
{\renewcommand{\arraystretch}{1.25}%
\begin{tabular}
{c@{\hspace{5mm}} c@{\hspace{5mm}} c@{\hspace{5mm}} c@{\hspace{5mm}} c}
\hline\hline
$M/m$ & This work & Ref.~\cite{RakDaiBlu12} & $M/m$ & This work\\
\hline
 0 & 3       &       &  6 & 2.01(2) \\
 1 & 3.19(1) & 3.155 &  7 & 1.77(1) \\
 2 & 3.05(1) &       &  8 & 1.56(3) \\
 3 & 2.85(1) &       &  9 & 1.19(4) \\
 4 & 2.56(1) &       & 10 & 0.99(1) \\
 5 & 2.31(1) &       & \\
\hline\hline
\end{tabular}}
\end{center}
\end{table}

Similar to the $(3+1)$ case, the bending in the $1^+$ energy results from
avoided crossing around $\alpha=1$ with another $1^+$ state (not shown).
The latter state has $\epsilon=4$ in the $\alpha=0$ limit and corresponds to the
$d$-wave atom-tetramer state. The same is true for the $2^+$ and $3^+$ states,
also shown in Fig.~\ref{Fig:Pentamer}, 
and indeed the energies of these state are close apart from the avoided
crossing region.

%~~~~~~~~~~~~~~~~~~~~~~~~~~~
\subsection {The (5+1) case}
%~~~~~~~~~~~~~~~~~~~~~~~~~~~

Adding another atom, we now move to the $(5+1)$ system.
Since no room is left in the $p$ shell, the additional 
atom can populate an excited $s$ shell, keeping the $0^-$ symmetry of
the $(4+1)$ core, or a $d$ shell, resulting in a $2^-$ state.

The energies of these states in a trap are tabulated in Table~\ref{tbl:Hexamer}
and plotted in Fig.~\ref{Fig:Hexamer}.

\begin{center}
\begin{figure}
\vskip 0 pt \includegraphics[clip,width=1\columnwidth]{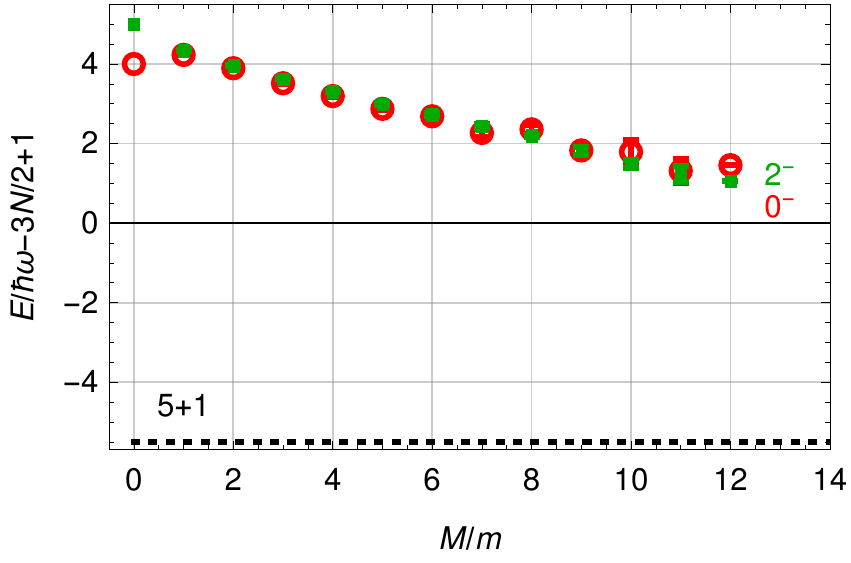}
\caption{
  The energy of the unitary $(5+1)$ trapped system is shown as a function of
  the mass ratio for a few lowest states.
  Symbols are the zero-range extrapolation from finite-range potentials.
  The Efimovian limit $s=0$ is the dotted horizontal line. For the mass ratios
  explored here, the scale factors do not cross this limit and therefore
  no $(5+1)$ Efimov effect exists. 
}
\label{Fig:Hexamer}
\end{figure}
\par\end{center}

\begin{table}
\begin{center}
  \caption{The energies of the two lowest $(5+1)$ hexamer states in a trap,
    with $0^-$ and $2^-$ symmetries, for various mass ratios.
    \label{tbl:Hexamer}}
\vspace{0.3cm}
{\renewcommand{\arraystretch}{1.25}%
\begin{tabular}
{c@{\hspace{5mm}} c@{\hspace{5mm}} c@{\hspace{5mm}} c@{\hspace{5mm}} c@{\hspace{5mm}} c}
\hline\hline
$M/m$ & $0^-$ & $2^-$ & $M/m$ & $0^-$ & $2^-$ \\
\hline
 0 & 4       & 5       &  6 & 2.7(1) & 2.73(4) \\
 1 & 4.23(1) & 4.34(1) &  7 & 2.3(1) & 2.44(6) \\
 2 & 3.89(3) & 3.96(2) &  8 & 2.4(1) & 2.20(3) \\
 3 & 3.52(3) & 3.63(2) &  9 & 1.8(1) & 1.8(1)  \\
 4 & 3.19(3) & 3.31(2) & 10 & 1.8(3) & 1.5(1)  \\
 5 & 2.87(4) & 2.99(3) & 11 & 1.3(3) & 1.2(2)  \\
\hline\hline
\end{tabular}}
\end{center}
\end{table}

As the mass ratio becomes larger, the $0^-$ and $2^-$ states becomes degenerate
within our error bars.

The Efimov limit corresponds here to $\epsilon=-5.5$.  
Our results show no sign for a $(5+1)$ Efimov state for any symmetry
up to $\alpha \le 12$.
As was have claimed, a different method would be probably needed to extend this
conclusion up the the $(4+1)$ Efimovian threshold.

%===================
\section{CONCLUSION}
%===================
We study mass-imbalanced mixtures of $N$ identical fermions interacting
resonantly with a distinguishable atom.
The scale factor, or the energy
of the unitary system in a harmonic trap, was calculated for a few lowest
states of the $N \le 5$ systems. We solve the trapped few-body system 
with finite-range inter-species potentials using the
stochastic variational method. The zero-range limit is then extrapolated.
The shell structure of the system is explored and the effect of level
repulsion is shown, revealing the significant change from the static-impurity
case to the dynamic-impurity case.
A series of Efimov states with $N=2,3$, and $4$
exist for large enough mass ratio. Nevertheless, no sign for
the existence of a $(5+1)$ Efimov effect is shown in the mass ratios
explored here, $\alpha\le 12$. Further studies that
would deal directly with the zero-range limit should be carried out to
check the validity of this statement for mass ratios up to the $(4+1)$
Efimovian threshold.

%========================
\section*{ACKNOWLEDGMENT}
%========================
I would like to thank Dmitry Petrov, Nir Barnea, Kalman Varga,
Johannes Kirscher, Ronen Weiss, and Yvan Castin
for useful discussions and communications.
This research was supported by the Pazi Fund.

%===========================================================================
\appendix*

%============================================
\section{Excited states in the $\alpha=0$ limit}
\label{excitedStates}
%============================================
For completeness, we list here the properties of the two lowest excited states
in the $\alpha=0$ limit.
The properties of the lowest-excited state are tabulated in
Table~\ref{tbl:ExcitedState}, while those of the next-to-lowest excited state
are tabulated in Table~\ref{tbl:2ExcitedState}.

\begin{table}
\begin{center}
  \caption{The lowest excited states properties for $\alpha=0$.
    Shown are the energy, the angular momentum, the parity, and the
    shell configuration for the $(N + 1)$ mixtures.
    \label{tbl:ExcitedState}}
\vspace{0.3cm}
{\renewcommand{\arraystretch}{1.25}%
\begin{tabular}
{c@{\hspace{5mm}} c@{\hspace{5mm}}  c@{\hspace{5mm}} c@{\hspace{5mm}} c}
\hline\hline 
System & $\epsilon$ & $\pi$ & $L$ & Configuration \\
\hline
1+1 & 2 & + & 0 & $1s$ \\
\hline %-----------------------
2+1 & 2 & + & 2 & $0s\,0d$ \\
\hline %-----------------------
\multirow{2}{*}{3+1} & \multirow{2}{*}{3} & + & 2 & $0s\,1s\,0d$ \\
\cline{3-5}
& & -- & 1,2,3 & $0s\,0p\,0d$ \\
\hline%-----------------------
\multirow{2}{*}{4+1} 
& \multirow{2}{*}{4} & + & 1,2,3 & $0s\,0p^2\,0d$ \\
\cline{3-5}
& & -- & 1,2,3 & $0s\,1s\,0p\,0d$ \\
\hline%-----------------------
\multirow{2}{*}{5+1} & \multirow{2}{*}{5}
& + & 1,2,3 & $0s\,1s\,0p^2\,0d$ \\
\cline{3-5}
& & -- & 2 & $0s\,0p^3\,0d$ \\
\hline\hline
\end{tabular}}
\end{center}
\end{table}

\begin{table}
\begin{center}
  \caption{The next-to-lowest excited state properties for $\alpha=0$.
    Shown are the energy, the angular momentum, the parity, and the shell
    configuration for the $(N + 1)$ mixtures.
    \label{tbl:2ExcitedState}}
\vspace{0.3cm}
{\renewcommand{\arraystretch}{1.25}%
\begin{tabular}
{c@{\hspace{5mm}} c@{\hspace{5mm}}  c@{\hspace{5mm}} c@{\hspace{5mm}} c}
\hline\hline 
System & $\epsilon$ & $\pi$ & $L$ & Configuration \\
\hline
1+1 & 4 & + & 0 & $2s$ \\
\hline %-----------------------
\multirow{4}{*}{2+1} &
\multirow{4}{*}{3} & + & 0 & $0s\,2s$ \\
\cline{3-5}
&  & \multirow{3}{*}{--} & 1 & $0s\,1p$ \\
&  &   & 1 & $1s\,0p$ \\
&  &   & 3 & $0s\,0f$ \\
\hline %-----------------------
\multirow{8}{*}{3+1} & 
\multirow{8}{*}{4} & \multirow{5}{*}{+} & 0,1,2 & $0s\,0p\,1p$ \\
&  &  & 1,3 & $0s\,0d^2$ \\
&  &  & 2,3,4 & $0s\,0p\,0f$ \\
&  &  & 0 & $0s\,1s\,2s$ \\
&  &  & 1 & $1s\,0p^2$ \\
\cline{3-5}
&  & \multirow{3}{*}{--} & 1 & $0s\,1s\,1p$ \\
&  &  & 1 & $0s\,2s\,0p$ \\
&  &  & 3 & $0s\,1s\,0f$ \\
\hline %-----------------------
\multirow{9}{*}{4+1} &
\multirow{9}{*}{5} & \multirow{4}{*}{+}
      & 0,1,2 & $0s\,1s\,0p\,1p$ \\
&  &  & 1 & $0s\,2s\,0p^2$ \\
&  &  & 1,3 & $0s\,1s\,0d^2$ \\
&  &  & 2,3,4 & $0s\,1s\,0p\,0f$ \\
\cline{3-5}
&  & \multirow{5}{*}{--}
     & 0,1,2,2,3,4 & $0s\,0p\,0d^2$ \\
&  &  & 0,1,2 & $0s\,0p^2\,1p$ \\
&  &  & 1 & $0s\,1s\,2s\,0p$ \\
&  &  & 2,3,4 & $0s\,0p^2\,0f$ \\
&  &  & 0 & $1s\,0p^3$ \\
\hline %-----------------------
\multirow{8}{*}{5+1} &
\multirow{8}{*}{6} & \multirow{4}{*}{+}
      & 0,1,2,2,3,4 & $0s\,0p^2\,0d^2$ \\
&  &  & 1 & $0s\,0p^3\,1p$ \\
&  &  & 1 & $0s\,1s\,2s\,0p^2$ \\
&  &  & 3 & $0s\,0p^3\,0f$ \\
\cline{3-5}
&  & \multirow{4}{*}{--}
      & 0 & $0s\,2s\,0p^3$ \\
&  &  & 0,1,2 & $0s\,1s\,0p^2\,1p$ \\
&  &  & 0,1,2,2,3,4 & $0s\,1s\,0p\,0d^2$ \\
&  &  & 2,3,4 & $0s\,1s\,0p^2\,0f$ \\
\hline\hline
\end{tabular}}
\end{center}
\end{table}

\end{document}